\documentclass[twocolumn,english,aps,prc,groupedaddress,amsmath,amssymb,showpacs]{revtex4}
\usepackage[T1]{fontenc}
\usepackage[latin9]{inputenc}
\usepackage{bm}
\usepackage{amsmath}
\usepackage{amssymb}

\makeatletter

\providecommand{\tabularnewline}{\\}

\@ifundefined{textcolor}{}
{%
 \definecolor{BLACK}{gray}{0}
 \definecolor{WHITE}{gray}{1}
 \definecolor{RED}{rgb}{1,0,0}
 \definecolor{GREEN}{rgb}{0,1,0}
 \definecolor{BLUE}{rgb}{0,0,1}
 \definecolor{CYAN}{cmyk}{1,0,0,0}
 \definecolor{MAGENTA}{cmyk}{0,1,0,0}
 \definecolor{YELLOW}{cmyk}{0,0,1,0}
 }

\usepackage{bm}\usepackage{dcolumn}\usepackage{babel}

\makeatother

\usepackage{babel}
\begin{document}

\title{Single-$\bm{j}$-shell studies of cross-conjugate nuclei and isomerism}

\author{L. Zamick}

\email{lzamick@physics.rutgers.edu}

\author{A. Escuderos}

\affiliation{Department of Physics and Astronomy, Rutgers University, New Brunswick,
NJ 08903, USA}
\begin{abstract}
Isomeric states for 4 nucleons with isospin $T=1$ are here considered.
A comparison is made of the lighter and heavier members of cross-conjugate
pairs. Although in the single $j$ shell the spectra in the two cases
should be identical, this is not the case experimentally. For the
former, the ground states all have angular momentum $J=2$. This result
is found in a single-$j$-shell calculation when the interaction is
obtained from the spectrum of two particles. In a single $j$ shell
($f_{7/2}$, $g_{9/2}$, $h_{11/2}$), the state with median angular
momentum $J\!=\!(J_{\text{max}}+1)/2$ is the ground state for the
heavier member of the pair provided one uses as an interaction the
spectrum of two holes. The ground state behaviour can also be explained
by rotational models. A new observation is that both in single-$j$
and in experiment the $J=2$ state in the heavier member and the $J=(J_{\text{max}}+1)/2$
state in the lighter member are isomeric. This criss-cross behaviour
shows that some remnants of the single-$j$-shell model persists and
indeed this model works surprisingly well. 
\end{abstract}


\maketitle

\section{Introduction}

In this work we develop a rule based on interesting behaviours of
nuclear spectra, or to be more precise spectra of four-nucleon states
with isospin $T\!=\!1$ in odd-odd nuclei. Such states consist of
either three protons and one neutron or three neutrons and one proton;
also three proton holes and one neutron hole or three neutron holes
and one proton hole. We find in single-$j$-shell calculations that
in the $f_{7/2}$, $g_{9/2}$, and $h_{11/2}$ shells, states with
median total angular momentum $J\!=\!(J_{\text{max}}+1)/2$ lie low
in energy and become isomeric for lighter members of cross-conjugate
pairs and ground states for the heavier members. Conversely, $J=2^{+}$
states are ground states for the lighter members and isomeric for
the heavier members. Although these calculations are relatively simple---not
large scale---, they are supported by experiment. We note that $J_{\text{max}}$
is equal to $M_{\text{max}}$. For three neutrons the maximum value
of $M$ is $j+(j-1)+(j-2)$ and for the single proton it is $j$.
Thus $J_{\text{max}}$ is equal to $(4j-3)$ whilst ($J_{\text{max}}$
+1)/2 is equal to $(2j-1)$. To briefly summarise the findings, we
note that for the three shells listed above the values of $J_{\text{max}}$
are 11, 15, and 19, respectively. Thus the $(J_{\text{max}}+1)/2$
rule gives values of 6, 8, and 10 for the low-lying isomeric (or ground)
states. We emphasize that the single-$j$-shell model is used only
to make qualitative statements about isomerism.

\section{The $\bm{f_{7/2}}$ shell}

We start with the $f_{7/2}$ shell where single-$j$-shell calculations
have already been performed and wave functions tabulated by Zamick,
Escuderos, and Bayman~\cite{escuderos0506050}; this reference is
based on previous work of Refs.~\cite{bmz63,mbz64,gf63}. The interaction
used consists of matrix elements taken from experiment---more precisely
from the spectrum of $^{42}$Sc and $^{42}$Ca (INTa). Zamick, however,
noted that for the upper half of the $f_{7/2}$ shell one obtains
better results by using matrix elements from the two-hole system $^{54}$Co
(INTb)~\cite{z02}. In single-$j$-shell calculations with both neutrons
and protons, we define the cross-conjugate of a given nucleus as one
in which protons are replaced by neutron holes and neutrons by proton
holes. Thus $^{52}$Fe is the cross-conjugate of $^{44}$Ti and $^{52}$Mn
is the cross-conjugate of $^{44}$Sc. If one uses the same charge
independant two-body interaction in both nuclei, the spectra for states
in this limited model space should be identical. In fact, although
the spectra are similar, they are not identical experimentally. The
10$^{+}$ state in $^{44}$Ti is below the 12$^{+}$, but in $^{52}$Fe
the reverse is true. In both cases the 12$^{+}$ state is isomeric
but the one in $^{52}$Fe has a much longer half-life because it cannot
decay to the 10$^{+}$ state. As seen in Table~\ref{tab:ti44} we
are successful in getting the 12$^{+}$ below the 10$^{+}$ by using
the spectrum of $^{54}$Co as input. The main difference in the two-body
spectra is that the $J=7^{+}$ state in $^{54}$Co is much lower in
energy than it is in $^{42}$Sc (see Table~\ref{tab:mes}).

\begin{table}[htb]
 \caption{\label{tab:ti44} Yrast spectra of $^{44}$Ti and $^{52}$Fe calculated
with the interactions INTa and INTb respectively (see text) and compared
with experiment~\cite{exp}.}

\begin{ruledtabular} %
\begin{tabular}{ccccc}
 & \multicolumn{4}{c}{$E$(MeV)}\tabularnewline
\hline 
 & \multicolumn{2}{c}{$^{44}$Ti} & \multicolumn{2}{c}{$^{52}$Fe}\tabularnewline
\hline 
$J$  & INTa  & Exp.  & INTb  & Exp. \tabularnewline
\hline 
0  & 0.000  & 0.000  & 0.000  & 0.000 \tabularnewline
1  & 5.669  &  & 5.442  & \tabularnewline
2  & 1.163  & 1.083  & 1.015  & 0.849 \tabularnewline
3  & 5.786  &  & 5.834  & \tabularnewline
4  & 2.790  & 2.454  & 2.628  & 2.384 \tabularnewline
5  & 5.871  &  & 6.463  & \tabularnewline
6  & 4.062  & 4.015  & 4.078  & 4.325 \tabularnewline
7  & 6.043  &  & 5.890  & \tabularnewline
8  & 6.084  & (6.509)  & 5.772  & 6.361 \tabularnewline
9  & 7.984  &  & 7.791  & \tabularnewline
10  & 7.384  & (7.671)  & 6.721  & 7.382 \tabularnewline
11  & 9.865  &  & 8.666  & \tabularnewline
12  & 7.702  & (8.040)  & 6.514  & 6.958 \tabularnewline
\end{tabular}\end{ruledtabular} 
\end{table}

Large space shell-model calculations for $^{52}$Fe were performed
by Ur \textit{et al.}~\cite{uetal98} using the KB3 interaction and
by Puddu~\cite{p09} using the GXPF1A interaction. Both groups get
a near degeneracy of $10_{1}^{+}$ and $12_{1}^{+}$ in $^{52}$Fe.
Thus, although they do not get $12^{+}$ sufficiently below $10^{+}$,
they do go on the right direction relative to $^{44}$Ti. Ur \textit{et
al.} attribute increased collectivity in $^{52}$Fe mainly to $p_{3/2}$
admixtures for the reason there are differences in the cross-conjugate
pairs.

We then examine the yrast spectrum of $^{44}$Sc calculated with the
interaction INTa (see Table~\ref{tab:sc44}). We consider two groups.
First for $J\!=\!6$, 5, 4, 3, 2, and 1, the energies in MeV are respectively
0.38, 1.28, 0.71, 0.76, 0.00, and 0.43 (the $J=0^{+}$ state has isospin
$T=2$ and is at an excitation energy of 3.047 MeV). We see that the
only state below the $J\!=\!6$ state is $J\!=\!2$. Thus, the lowest
multipolarity for decay is $E4$ and so the $J\!=\!6$ state is calculated
to be isomeric. For the second group with $J\!=\!11$, 10, 9, 8, and
7, the energies in MeV are respectively 4.64, 4.79, 3.39, 3.10, and
1.27. The $J^{\pi}\!=\!11^{+}$ state can decay via an $E2$ transition
to the $J^{\pi}\!=\!9^{+}$ state so it should not be isomeric.

\begin{table}[htb]
 \caption{\label{tab:sc44} Yrast spectra of $^{44}$Sc and $^{52}$Mn calculated
with the interactions INTa and INTb respectively (see text) and compared
with experiment~\cite{exp}.}

\begin{ruledtabular} %
\begin{tabular}{ccccc}
 & \multicolumn{4}{c}{$E$(MeV)}\tabularnewline
\hline 
 & \multicolumn{2}{c}{$^{44}$Sc} & \multicolumn{2}{c}{$^{52}$Mn}\tabularnewline
\hline 
$J$  & INTa  & Exp.  & INTb  & Exp. \tabularnewline
\hline 
0  & 3.047  &  & 2.774  & \tabularnewline
1  & 0.432  & 0.667  & 0.443  & 0.546 \tabularnewline
2  & 0.000  & 0.000  & 0.202  & 0.378 \tabularnewline
3  & 0.764  & 0.762  & 0.836  & 0.825 \tabularnewline
4  & 0.713  & 0.350  & 0.851  & 0.732 \tabularnewline
5  & 1.276  & 1.513  & 1.404  & 1.254 \tabularnewline
6  & 0.381  & 0.271  & 0.000  & 0.000 \tabularnewline
7  & 1.272  & 0.968  & 1.819  & 0.870 \tabularnewline
8  & 3.097  &  & 2.572  & (2.286) \tabularnewline
9  & 3.390  & 2.672  & 2.792  & (2.908) \tabularnewline
10  & 4.793  & 4.114  & 4.365  & 4.164 \tabularnewline
11  & 4.638  & 3.567  & 3.667  & (3.837) \tabularnewline
\end{tabular}\end{ruledtabular} 
\end{table}

We now look at experiment. In $^{44}$Sc the lowest $J^{\pi}\!=\!6^{+}$
state has a half-life of 58.6 hours---it is indeed isomeric.

But we should also consider the cross-conjugate nucleus $^{52}$Mn
consisting of three proton holes and one neutron hole relative to
$^{56}$Ni. We see that here the $J^{\pi}\!=\!6^{+}$ state is the
ground state with a half-life of 5.591 days. As mentioned before,
if we use the same interaction here as we did for $^{44}$Sc, we would
not get the $J=6^{+}$ state as the ground state. But as seen in Table~\ref{tab:sc44},
when we use as input the spectrum of the two-hole system $^{54}$Co,
we do get $J=6^{+}$ as the ground state.

There is some indication that in heavier nuclei the state with $J\!=\! J_{\text{max}}$
should be isomeric. However, the $J^{\pi}\!=\!11^{+}$ state at 3.57
MeV in $^{44}$Sc has a half-life of 48 ps whilst the corresponding
$J^{\pi}\!=\!11^{+}$ state in $^{52}$Mn at 3.84 MeV has a half-life
of 15.1 ps.

We could not find large space shell-model calculations of $^{52}$Mn
in the literature, but there is a single-$j$-shell calculation in
the work of Avrigeanu \textit{et al.}~\cite{aetal76}. This accompanies
their experimental work on high-spin states in this nucleus.

\section{The $\bm{g_{9/2}}$ shell}

In previous work~\cite{PhysRevC.73.044302}, calculations were performed
in the $g_{9/2}$ shell where the emphasis was on partial dynamical
symmetries. However, many residual calculations were lying around
which had not been carefully examined. Nara Singh~\cite{nara2011}
reported the finding by his group of a $J^{\pi}\!=\!16^{+}$ isomeric
state in $^{96}$Cd that beta decayed to a $J^{\pi}\!=\!15^{+}$ state
in $^{96}$Ag, which is also isomeric. This largely stimulated the
work done here on isomerism. We also note a combination of experiment
and shell-model calculations by K. Schmidt \textit{et al.}~\cite{Schmidt1997185}
and L. Batist \textit{et al.}~\cite{Batist2003245}. The topics addressed
in these works are decay properties of very neutron-deficient isotopes
of silver and cadmium, as well as isomerism in $^{96}$Ag.

We show results for two interactions: INTc and INTd. The $T=1$ matrix
elements are obtained from the spectrum of $^{98}$Cd, that is, two
proton holes. Unfortunately, the spectrum of $^{98}$In is not known,
so we cannot get the $T=0$ matrix elements from experiment. We use
a delta interaction to generate the $T=0$ matrix elements for INTc.
Noting that in the $f_{7/2}$ shell the state with $J=J_{\text{max}}$,
i.e. $J=7$, comes much lower for two holes than it does for two particles,
we simulate this behaviour in INTd in the $g_{9/2}$ shell by changing
the $J_{\text{max}}=9$ energy from 1.4964 MeV to 0.7500 MeV, leaving
all other two-body matrix elements the same. This interaction should
be more appropriate for the four-hole system.

\begin{table}[htb]
 \caption{\label{tab:g92} Energy levels for the case of 3 protons and 1 neutron
in the $g_{9/2}$ shell with the interactions INTc and INTd (see text),
and compared with the experimental data for $^{96}$Ag.}

\begin{ruledtabular} %
\begin{tabular}{cccc}
 & \multicolumn{3}{c}{$E$(MeV)}\tabularnewline
\hline 
$J$  & INTc  & INTd  & Exp. \tabularnewline
\hline 
0  & 0.246  & 0.900  & \tabularnewline
1  & 0.463  & 0.483  & \tabularnewline
2  & 0.000  & 0.097  & \tabularnewline
3  & 0.638  & 0.588  & \tabularnewline
4  & 0.394  & 0.349  & \tabularnewline
5  & 0.774  & 0.737  & \tabularnewline
6  & 0.450  & 0.371  & \tabularnewline
7  & 0.850  & 0.861  & \tabularnewline
8  & 0.350  & 0.000  & 0.000 \tabularnewline
9  & 0.872  & 0.492  & 0.470 \tabularnewline
10  & 2.188  & 1.748  & (1.719) \tabularnewline
11  & 2.344  & 1.930  & (1.976) \tabularnewline
12  & 3.004  & 2.550  & \tabularnewline
13  & 3.087  & 2.556  & 2.643 \tabularnewline
14  & 3.382  & 3.070  & \tabularnewline
15  & 3.287  & 2.645  & 2.643+$x$ \tabularnewline
\end{tabular}\end{ruledtabular} 
\end{table}

With the INTc interaction, the $J=2^{+}$ state is the ground state
and should be long lived. The $J=8^{+}$ is at an excitation energy
of 0.350 MeV, so only the $J=0^{+}$ ($T=2$) and $J=2^{+}$ states
are below it. So this state should be isomeric. But for INTd, where
we lowered the energy of the $J=9^{+}$ two-body matrix element, the
$J=8^{+}$ state is now the ground state and is of course long lived.
The $J=2^{+}$ state is very low lying (0.097 MeV) and is isomeric.
At high spin with INTd the $J=15^{+}$ state is at 2.645 MeV while
the $J=13^{+}$ state is at 2.556 MeV. Because they are so close in
energy, the $J=15^{+}$ state is isomeric.

Concerning experiment in Refs.~\cite{Batist2003245,Schmidt1997185},
nearly degenerate $J^{\pi}\!=\!2^{+}$ and $J^{\pi}\!=\!8^{+}$ isomers
are shown with respective half-lives of 6.9(6)~s and 4.40(6)~s.
We see that also in this shell the $(J_{\text{max}}+1)/2$ rule is
verified.

We find that, unlike in the $f_{7/2}$ shell, here in $g_{9/2}$ our
calculation with INTd leads to an isomeric state for $J\!=\! J_{\text{max}}=15$
and this supports the experimental findings of Nara Singh \cite{nara2011}.
We now refer to the experimental works of Grzywacz \textit{et al.}~\cite{Phys.Rev.C.55.1126}
and Grawe \textit{et al.}~\cite{Eur.Phys.J.A.27.257}. The latter
work also includes large-scale shell-model calculations and points
out that there are many spin-gap states in the $^{100}$Sn region.
A near degeneracy of the two states in $^{96}$Ag is shown in Fig.~1
of Grawe \textit{et al.}, however with the $J\!=\!13^{+}$ state ever
so slightly below the $J\!=\!15^{+}$ state.

\section{The $\bm{h_{11/2}}$ shell}

We include here results for the $h_{11/2}$ shell. A $Q\cdot Q$ interaction
is used with strength such that the $J\!=\!2\!-\! J\!=\!0$ splitting
for two nucleons is 0.244 MeV. No comparison with experiment can be
made, but we want to show that the pattern of behavior of the previous
shells persists here as well. The energies in MeV for the first group
with $J\!=\!10$, 9, 8, 7, 6, 5, 4, 3, 2, and 1 are respectively 0.21,
1.30, 1.58, 1.33, 1.03, 0.75, 0.49, 0.28, 0.11, and 0.00. Only states
with $J\!=\!2$ and 1 lie below the $J\!=\!10$ state in this calculation.
This fits in very nicely with the $(J_{\text{max}}+1)/2$ rule. For
the second group with angular momenta $J\!=\!19$, 18, 17, 16, 15,
14, 13, 12, and 11, the energies in MeV are respectively 2.43, 3.11,
2.85, 3.08, 2.66, 2.47, 1.97, 1.46, and 0.85. The highest-spin state
below the $J^{\pi}\!=\!19^{+}$ state in this calculation has $J^{\pi}\!=\!13^{+}$.
Thus the $J^{\pi}\!=\!19^{+}$ state is predicted to be isomeric.

\section{Lighter Nuclei}

The single-shell model is not expected to work for light nuclei and
no calculation will be attempted. Still it is of interest to show
the systematics of these nuclei in the $d_{5/2}$ and $p_{3/2}$ shells.

The nucleus $^{20}$F is a special case. Here $(J_{\text{max}}+1)/2$
is equal to 4. Clearly this state can readily decay to the lower $J^{\pi}\!=\!2^{+}$
state. Experimentally the lowest 3 states are: $J^{\pi}\!=\!2^{+}$
(ground), $J^{\pi}\!=\!3^{+}$ at 0.656 MeV, and $J^{\pi}\!=\!4^{+}$
at 0.823 MeV. It is understandable that when ($J_{\text{max}}+1)/2$
differs from the ground state spin by only 2 units the isomerism is
no longer present. However, despite two decay channels being open,
the half-life of the 4$^{+}$ state is surprisingly long at 55 ps.
The next longest half-life listed is 1.36~ps for the $J^{\pi}\!=\!1^{-}$
state at 0.984 MeV. In the cross-conjugate nucleus $^{24}$Na the
$J=4^{+}$ state is the ground state and is isomeric with a lifetime
of 14.997~h. There is an excited $1^{+}$ state at 0.472 MeV with
a half-life of 20.18 ms and a $2^{+}$ state at 0.563 MeV with a half-life
of 36 ps. Other transitions in this nucleus are much shorter.

For $^{8}$Li the value of ($J_{\text{max}}+1)/2$) is 2. This corresponds
to the ground state which is of course isomeric with a half-life of
839.9 ms. This nucleus is its own cross-conjugate.

\section{Explanations of the isomerisms}

We have admittedly done some very simple calculations, but that is
the point. One should do such calculations to search for interesting
behaviors. Later one can supplement these with more detailed calculations.
The simple calculations are useful when effects are large as in the
case of the $(J_{\text{max}}+1)/2$ rule.

A key to understanding the isomerisms comes from the works of Gallagher-Moszkowski~\cite{PhysRev.111.1282}.
They developed a scheme for obtaining and predicting the ground state
spins of odd-odd nuclei. Briefly stated, the value of the total angular
momentum is predicted to be ($\Omega_{p}+\Omega_{n})$ where $\Omega$
is the component of the angular momentum along the symmetry axis.
We can make a connection with this by noting that for all the heavier
members of the cross-conjugate pairs in the previous sections, the
ground states have $J$ values equal to ($J_{\text{max}}+1)/2$, which
is the same as ($\Omega_{n}+\Omega_{p}$). In more detail the lighter
members of the cross-conjugate pairs have one proton with $\Omega_{p}=1/2$
and three neutrons with $\Omega_{n}=3/2$. This leads to a ground
state spin $J=2$ which is verified experimentally for all nuclei
here considered with the qualification that the spin is not yet known
for $^{84}$Nb---in the tables, three possibilities are listed: ($1^{+}$,
$2^{+}$, $3^{+}$). To form a cross-conjugate nucleus, we replace
a proton by a neutron-hole and a neutron by a proton-hole. Thus one
proton becomes $2j$ neutrons and three neutrons becomes ($2j-2$)
protons. The vaule of $\Omega$ is then $j+(j-1)=(2j-1)$, and this
is also ($J_{\text{max}}+1)/2$.

To complete the argument, we note that in the single-$j$-shell model
a nucleus and its cross-conjugate partner should have identical spectra.
This is not the case experimentally. The lighter members have $J=2$
ground states and the heavier ones $J=(2j-1)$ ground states. As far
as the isomerism rule is concerned, we would argue that for the lighter
members of the cross-congugate pairs the shell effects are present,
which, although not srong enough to maintain identical spectra with
their partners, are nevertheless strong enough to keep the ($J_{\text{max}}+1)/2$
states sufficiently low as to be isomeric in the lighter members and
the $J=2^{+}$ states to be isomeric in the heavier ones.

\section{Isobaric analog states---$\bm{f_{7/2}}$ vs. $\bm{g_{9/2}}$}

The $J=0^{+}$ states in Tables~\ref{tab:sc44} and \ref{tab:g92}
have isospins $T=2$ while the other states have $T=1$. The $J=0^{+}$
states in $^{96}$Ag are isobaric analog states of $J=0^{+}$ states
of the four proton-hole nucleus $^{96}$Pd. Note that with the interactions
that we have used, the $J=0^{+}$ states lie much lower in the $g_{9/2}$
shell than in the $f_{7/2}$ shell, as far as a system of three protons
and one neutron is concerned. There actually are two $T=2$, $J=0^{+}$
states for $(g_{9/2})^{4}$, only one for $f_{7/2}$. With INTd the
lowest $J=0^{+}$ state is at an excitation of 0.900 MeV, a prediction
for $^{96}$Ag. In $^{44}$Sc and $^{52}$Mn the excitation energies
are 3.047 and 2.774 MeV respectively. Some caution must be used because
of the uncertainty of the $T=0$ two-body matrix elements in the $g_{9/2}$
shell.

\section{A brief discussion of high-spin states in $^{\bm{96}}\text{Cd}$}

Three very closely timed publications have appeared on the subject
of isomerism for $A=96$. In reference~\cite{nara2011} Nara Singh
\textit{et al.} first found a $J=16^{+}$ isomeric state in $^{96}$Cd.
Indeed at the time of this writing this is the only known state in
this nucleus. A recent work by A.D. Becerril \textit{et al.}~\cite{b84}
is very relevant to the work discussed here. They find two isomeric
states in $^{96}$Ag. They do not assign spins but they are probably
15$^{+}$ and 13$^{-}$ . Then there is the work of P. Boutachkov
\textit{et al.}~\cite{bo84} which follows from the findings of reference~\cite{nara2011}.
They observe the direct decay of the isomeric 16$^{+}$ state of $^{96}$Cd
to the 15$^{+}$ isomeric state in $^{96}$Ag and are able to determine
the spins of this and other isomers.

Our single-$j$-shell calculation also yields a $J=16^{+}$ isomer
for $^{96}$Cd (see Table~\ref{tab:96cd}). We see that the $J=16^{+}$
state is calculated to be lower than $J=15^{+}$ or $14^{+}$ for
both interactions. This guarantees isomerism in this model space.
In principle this could be upset by the appearance of negative parity
states and electric dipole transitions but this does not seem to be
the case experimentally.

\begin{table}[htb]
 \caption{\label{tab:96cd} Calculated energies of states for $^{96}$Cd from
$J=10^{+}$ to $16^{+}$.}

\begin{ruledtabular} %
\begin{tabular}{ccc}
 & \multicolumn{2}{c}{$E$(MeV)}\tabularnewline
\hline 
$J^{\pi}$  & INTc  & INTd \tabularnewline
\hline 
$10^{+}$  & 4.570  & 4.617 \tabularnewline
$11^{+}$  & 5.312  & 5.564 \tabularnewline
$12^{+}$  & 5.232  & 5.630 \tabularnewline
$13^{+}$  & 5.696  & 5.895 \tabularnewline
$14^{+}$  & 5.430  & 5.030 \tabularnewline
$15^{+}$  & 6.625  & 5.564 \tabularnewline
$16^{+}$  & 5.506  & 4.937 \tabularnewline
\end{tabular}\end{ruledtabular} 
\end{table}

\section{Symmetries in the $\bm{g_{9/2}}$ shell}

One good reason to study the single-$j$-shell limit is that certain
symmetries appear which can serve as focal points as we go to larger
spaces. For example, it was noted in reference~\cite{PhysRevC.73.044302}
that in the case of four identical nucleons in the $g_{9/2}$ shell
with total angular momentum $J=4$ there is a special eigenstate which
emerges no matter what interaction is used. This is a seniority $v=4$
state that does not mix with the $v=2$ state and more surprisingly
does not mix with the other $v=4$ state (three $J=4$ states in all).
The analog of this state in $^{96}$Pd appears in the spectra of $^{96}$Ag
and $^{96}$Cd. Usually the components of the wave function appear
the same no matter what interaction is used. The exception is if there
are degeneracies. It was already pointed out in reference~\cite{PhysRevC.73.044302}
that with a pairing interaction the two $v=4$ states are degenerate
and so any linear combinations can emerge from a matrix diagonalization---therefore
the uniqueness of the special $v=4$ state is lost. Too much symmetry
is a bad thing. In $^{96}$Cd another strange behaviour arises. With
a $Q\cdot Q$ interaction there is a degeneracy of this $T=2$ special
state with a $T=0$ state. One then encounters a technical problem:
the computer program puts forth some arbitrary linear combination
of these two states. One can get rid of this problem by adding a $t(1)\cdot t(2)$
interaction to $Q\cdot Q$. This will separate the two states and
the $T=2$ unique wave function will look the same as for any other
interaction.

Again in $^{96}$Cd it was pointed out in reference~\cite{zr84}
that if one used a two-body interaction for which the two-body matrix
elements with isospin $T=0$ (i.e. the odd-$J$ ones) were set to
zero or to a constant, then the $J=16^{+}$ state would be degenerate
with a $J=14^{+}$ state both having the configuration {[}$J_{p}=8$,
$J_{n}=8${]}. The wave functions and energies are shown for INTd
and INTd with $T=0$ matrix elements set equal to zero in Tables~\ref{tab:96cd-intd}
and \ref{tab:96cd-intd0} respectively. With the latter interaction,
we see that for $J^{\pi}=14^{+}$ the states have good dual quantum
numbers {[}$J_{p},J_{n}${]}. With the INTd interaction the $J=16^{+}$
state is at 4.937 MeV. There are three $J=14^{+}$ states, one at
5.110 MeV, a second with mostly an {[}8, 8{]} component at 6.498 MeV---these
have isospin $T=0$---, and a $T=1$ state at 6.703 MeV. In other
words, the $J^{\pi}=14^{+}$ state at 6.498~MeV is a pure {[}8,~8{]}
configuration and the others are {[}6,~8{]}$\pm${[}8,~6{]}. For
the full INTd interaction, the wave functions are more complicated.
Note also that in Table~\ref{tab:96cd-intd0} the $J^{\pi}=14^{+}$
state with configuration {[}8,~8{]} is degenerate with the $J^{\pi}=16^{+}$
state, also with {[}8,~8{]} configuration; both states have isospin
$T=0$. We see that the $T=0$ two-body interaction is responsible
for removing the 14--16 degeneracy for {[}8, 8{]}.

Note futher in Table 6 a three fold degeneracy.Staes with angular
momenta 11,13 and 14 all at 5.3798 MeV. Further more these states
heve the same dual quantum numbers (6,8) and (8,6). We have here a
partial dynamical symmetry. The T=0 states that belong to this symmetry
have totoal angular momenta that cannot occur for systems of indentical
nucleons i.e. thy cannot osscur in the single j shell in $^{96}$Pd,
a system of 4 neutron holes . As a counterpoint we include the wave
functions for J=12. There is no symmetry here--the wave functions
have components with various values of (J$_{p}$,J$_{n}$).The symmetry
displays iteslf with states with J=11.13.14. and 16--these cannot
occur for 4 identical particles. Now J=15 also does not occur for
4 identical particles. Howver this state has isospin T=1. The symmetry
only occurs for T=0 states.

\begin{table}[htb]
 \caption{\label{tab:96cd-intd} Wave functions and energies (in MeV, at the
top) of selected states of $^{96}$Cd calculated with the INTd interaction.}

\begin{ruledtabular} %
\begin{tabular}{cccccccc}
\multicolumn{2}{l}{$J=11$} &  &  &  &  &  & \tabularnewline
 &  & 5.5640  & 5.6482  & 6.4693  & 6.6384  & 6.9319  & 8.1822 \tabularnewline
$J_{p}$  & $J_{n}$  &  &  & $T=1$  & $T=1$  & $T=1$  & $T=1$ \tabularnewline
4  & 8  & 0.4709  & $-0.6359$  & $-0.2463$  & 0.3092  & $-0.4544$  & 0.1051 \tabularnewline
6  & 6  & 0.2229  & 0.0000  & 0.8712  & 0.0000  & $-0.3121$  & $-0.3065$ \tabularnewline
6  & 8  & 0.4607  & $-0.3092$  & $-0.0631$  & $-0.6359$  & 0.4432  & $-0.2956$ \tabularnewline
8  & 4  & 0.4709  & 0.6359  & $-0.2463$  & $-0.3092$  & $-0.4544$  & 0.1051 \tabularnewline
8  & 6  & 0.4607  & 0.3092  & $-0.0631$  & 0.6359  & 0.4432  & $-0.2956$ \tabularnewline
8  & 8  & 0.2869  & 0.0000  & 0.3343  & 0.0000  & 0.3110  & 0.8421 \tabularnewline
\hline 
\multicolumn{2}{l}{$J=12$} &  &  &  &  &  & \tabularnewline
 &  & 5.0303  & 5.8274  & 6.1835  & 6.7289  & 6.8648  & 9.0079 \tabularnewline
$J_{p}$  & $J_{n}$  &  &  &  & $T=1$  & $T=1$  & $T=2$ \tabularnewline
4  & 8  & 0.4364  & 0.3052  & $-0.3894$  & $-0.3592$  & $-0.5903$  & 0.2957 \tabularnewline
6  & 6  & 0.7797  & $-0.4079$  & 0.0000  & 0.2927  & 0.0000  & $-0.3742$ \tabularnewline
6  & 8  & 0.0344  & 0.5602  & $-0.5903$  & 0.2078  & 0.3894  & $-0.3766$ \tabularnewline
8  & 4  & 0.4364  & 0.3052  & 0.3894  & $-0.3592$  & 0.5903  & 0.2957 \tabularnewline
8  & 6  & 0.0344  & 0.5602  & 0.5903  & 0.2078  & $-0.3894$  & $-0.3766$ \tabularnewline
8  & 8  & 0.0940  & 0.1402  & 0.0000  & 0.7550  & 0.0000  & 0.6337 \tabularnewline
\hline 
\multicolumn{2}{l}{$J=13$} &  &  &  &  &  & \tabularnewline
 &  & 5.8951  & 6.1898  & 7.5023  &  &  & \tabularnewline
$J_{p}$  & $J_{n}$  &  & $T=1$  & $T=1$  &  &  & \tabularnewline
6  & 8  & 0.7071  & 0.6097  & $-0.3581$  &  &  & \tabularnewline
8  & 6  & $-0.7071$  & 0.6097  & $-0.3581$  &  &  & \tabularnewline
8  & 8  & 0.0000  & 0.5065  & 0.8623  &  &  & \tabularnewline
\hline 
\multicolumn{2}{l}{$J=14$} &  &  &  &  &  & \tabularnewline
 &  & 5.1098  & 6.4980  & 6.7036  &  &  & \tabularnewline
$J_{p}$  & $J_{n}$  &  &  & $T=1$  &  &  & \tabularnewline
6  & 8  & 0.6943  & $-0.1339$  & $-0.7071$  &  &  & \tabularnewline
8  & 6  & 0.6943  & $-0.1339$  & 0.7071  &  &  & \tabularnewline
8  & 8  & 0.1894  & 0.9819  & 0.0000  &  &  & \tabularnewline
\hline 
\multicolumn{2}{l}{$J=15$} &  &  &  &  &  & \tabularnewline
 &  & 6.2789  &  &  &  &  & \tabularnewline
$J_{p}$  & $J_{n}$  & $T=1$  &  &  &  &  & \tabularnewline
8  & 8  & 1.0000  &  &  &  &  & \tabularnewline
\hline 
\multicolumn{2}{l}{$J=16$} &  &  &  &  &  & \tabularnewline
 &  & 4.9371  &  &  &  &  & \tabularnewline
$J_{p}$  & $J_{n}$  &  &  &  &  &  & \tabularnewline
8  & 8  & 1.0000  &  &  &  &  & \tabularnewline
\end{tabular}\end{ruledtabular} 
\end{table}

\begin{table}[htb]
 \caption{\label{tab:96cd-intd0} Wave functions and energies (in MeV, at the
top) of selected states of $^{96}$Cd calculated with the interaction
INTd with $T=0$ matrix elements set to zero.}

\begin{ruledtabular} %
\begin{tabular}{cccccccc}
\multicolumn{2}{l}{$J=11$} &  &  &  &  &  & \tabularnewline
 &  & 5.0829  & 5.3798  & 6.8295  & 7.4699  & 7.5178  & 7.8842 \tabularnewline
$J_{p}$  & $J_{n}$  &  &  & $T=1$  & $T=1$  & $T=1$  & $T=1$ \tabularnewline
4  & 8  & 0.7071  & 0.0000  & 0.2933  & $-0.5491$  & 0.3351  & $-0.0121$ \tabularnewline
6  & 6  & 0.0000  & 0.0000  & 0.2913  & 0.5605  & 0.6482  & $-0.4253$ \tabularnewline
6  & 8  & 0.0000  & 0.7071  & 0.5350  & 0.0396  & $-0.4111$  & $-0.2079$ \tabularnewline
8  & 4  & $-0.7071$  & 0.0000  & 0.2933  & $-0.5491$  & 0.3351  & $-0.0121$ \tabularnewline
8  & 6  & 0.0000  & $-0.7071$  & 0.5350  & 0.0396  & $-0.4111$  & $-0.2079$ \tabularnewline
8  & 8  & 0.0000  & 0.0000  & 0.4130  & 0.2822  & 0.1319  & 0.8558 \tabularnewline
\hline 
\multicolumn{2}{l}{$J=12$} &  &  &  &  &  & \tabularnewline
 &  & 5.1165  & 5.2336  & 5.4865  & 7.5293  & 7.5959  & 12.4531 \tabularnewline
$J_{p}$  & $J_{n}$  &  &  &  & $T=1$  & $T=1$  & $T=2$ \tabularnewline
4  & 8  & 0.5699  & 0.2803  & $-0.0961$  & $-0.4783$  & 0.5208  & 0.2957 \tabularnewline
6  & 6  & 0.5712  & $-0.7151$  & 0.1498  & 0.0000  & 0.0000  & $-0.3742$ \tabularnewline
6  & 8  & 0.0925  & 0.3679  & 0.4629  & $-0.5208$  & $-0.4783$  & $-0.3766$ \tabularnewline
8  & 4  & 0.5699  & 0.2803  & $-0.0961$  & 0.4783  & $-0.5208$  & 0.2957 \tabularnewline
8  & 6  & 0.0925  & 0.3679  & 0.4629  & 0.5208  & 0.4783  & $-0.3766$ \tabularnewline
8  & 8  & $-0.0846$  & $-0.2465$  & 0.7284  & 0.0000  & 0.0000  & 0.6337 \tabularnewline
\hline 
\multicolumn{2}{l}{$J=13$} &  &  &  &  &  & \tabularnewline
 &  & 5.3798  & 7.6143  & 7.8873  &  &  & \tabularnewline
$J_{p}$  & $J_{n}$  &  & $T=1$  & $T=1$  &  &  & \tabularnewline
6  & 8  & 0.7071  & 0.5265  & $-0.4721$  &  &  & \tabularnewline
8  & 6  & $-0.7071$  & 0.5265  & $-0.4721$  &  &  & \tabularnewline
8  & 8  & 0.0000  & 0.6676  & 0.7445  &  &  & \tabularnewline
\hline 
\multicolumn{2}{l}{$J=14$} &  &  &  &  &  & \tabularnewline
 &  & 5.3798  & 5.6007  & 7.8515  &  &  & \tabularnewline
$J_{p}$  & $J_{n}$  &  &  & $T=1$  &  &  & \tabularnewline
6  & 8  & 0.7071  & 0.0000  & $-0.7071$  &  &  & \tabularnewline
8  & 6  & 0.7071  & 0.0000  & 0.7071  &  &  & \tabularnewline
8  & 8  & 0.0000  & 1.0000  & 0.0000  &  &  & \tabularnewline
\hline 
\multicolumn{2}{l}{$J=15$} &  &  &  &  &  & \tabularnewline
 &  & 7.9251  &  &  &  &  & \tabularnewline
$J_{p}$  & $J_{n}$  & $T=1$  &  &  &  &  & \tabularnewline
8  & 8  & 1.0000  &  &  &  &  & \tabularnewline
\hline 
\multicolumn{2}{l}{$J=16$} &  &  &  &  &  & \tabularnewline
 &  & 5.6007  &  &  &  &  & \tabularnewline
$J_{p}$  & $J_{n}$  &  &  &  &  &  & \tabularnewline
8  & 8  & 1.0000  &  &  &  &  & \tabularnewline
\end{tabular}\end{ruledtabular} 
\end{table}

\section{Added comments}

We emphasize here that we are making only qualitative statements about
isomerism, i.e. which angular momenta are and are not isomeric. We
make comparisons of cross-conjugate pairs. Cross-conjugation is a
single-$j$-shell concept and so we invoke this model for insight
into the behaviour of these four-nucleon $T=1$ systems. We then note
that memory of the single $j$ shell persists even in larger space
calculations and indeed in nature. This explains the criss-cross behaviour
so that $J=2^{+}$ states in lighter members of a cross-conjugate
pair are ground states and in the heavier members they are sufficiently
low lying so as to be isomeric. Likewise ($J_{\text{max}}+1)/2$ states
are sufficiently low lying so as to be isomeric for lighter members
and ground states for heavier members. An important point in obtaining
these results is that one should use as input the two-particle spectrum
for the lighter member of the cross-conjugate pair and the two-hole
spectrum for the heavier pair. The most obvious difference is that
the energy of the two-hole state with $J=J_{\text{max}}=2j$ is much
lower than the corresponding energy for two particles.

It should be further noted that the energy levels come out fairly
well in the single-$j$-shell model when compared with experiment
(see Tables~\ref{tab:ti44}, \ref{tab:sc44}, and \ref{tab:g92}).
Note that the sudden drop in the $J^{\pi}=9^{+}$ energy of $^{96}$Ag
is correctly reproduced. This shows that the single-$j$-shell model
has considerable validity for the cases considered.

Most importantly we feel that after addressing the properties of a
given nucleus, either theoretically or experimentally, one should
try to see if the specific results are part of a bigger picture. This
is certainly the case here. For example, the striking analogous behaviours
in $^{52}$Mn and $^{96}$Ag lead us to conclude that both $J=2^{+}$
and $(J_{\text{max}}+1)/2$ states should be long-lived.
\begin{acknowledgments}
One author (L.Z.) benefited from attending the Weizmann post-NPA5
workshop. He was supported as a visiting professor at the Weizmann
Institute by a Morris Belkin award. He thanks Diego Torres for his
help in preparing this manuscript and for his critical suggestions. 
\end{acknowledgments}
\appendix

\section{Interactions discussed in this work}

We first show in Table~\ref{tab:mes} the two-body matrix elements
that we used in this work in increasing spin from $J=0$ to $J=J_{\text{max}}$.
The even spins have isospin $T=1$ and the odd ones $T=0$.

We next consider the large scale interactions. In Ref.~\cite{uetal98}
the KB3 interaction was used in a complete $f$-$p$ space ($f_{7/2}$,
$p_{3/2}$, $f_{5/2}$, $p_{1/2}$) for the study of $^{52}$Fe; in
Ref.~\cite{p09} the GXPF1A interaction was used for the same nucleus
and model space. In Ref.~\cite{Batist2003245}, to study mainly $^{96}$Pd,
the SLG and F-FIT interactions were used in the model space ($p_{1/2}$,
$g_{9/2}$), together with the JS interaction in a somewhat larger
model space (allowing single-nucleon excitations to the orbitals $g_{7/2}$,
$d_{5/2}$, $s_{1/2}$, $d_{3/2}$). Again the model space ($p_{1/2}$,
$g_{9/2}$) was used in Ref.~\cite{b84} for $^{96}$Ag with the
SLGT interaction, while the jj44b interaction was also used but within
the model space ($p_{3/2}$, $f_{5/2}$, $p_{1/2}$, $g_{9/2}$).
Finally, in Ref.~\cite{bo84} various interactions were used: GF
in the space ($p_{1/2}$, $g_{9/2}$), FPG in ($p_{3/2}$, $f_{5/2}$,
$p_{1/2}$, $g_{9/2}$), and GDS in ($g_{9/2}$, $g_{7/2}$, $d_{5/2}$,
$s_{1/2}$, $d_{3/2}$).

\begin{table}[htb]
 \caption{\label{tab:mes} Two-body matrix elements in increasing spin from
$J=0$ to $J=J_{\text{max}}$. The even spins have isospin $T=1$
and the odd ones $T=0$.}

\begin{ruledtabular} %
\begin{tabular}{cccccc}
 & \multicolumn{2}{c}{$f_{7/2}$} & \multicolumn{2}{c}{$g_{9/2}$} & $h_{11/2}$ \tabularnewline
\hline 
$J$  & INTa  & INTb  & INTc  & INTd  & $Q\cdot Q$\tabularnewline
\hline 
0  & 0.0000  & 0.0000  & 0.0000  & 0.0000  & $-1.0000$ \tabularnewline
1  & 0.6111  & 0.5723  & 1.1387  & 1.1387  & $-0.9161$ \tabularnewline
2  & 1.5863  & 1.4465  & 1.3947  & 1.3947  & $-0.7544$ \tabularnewline
3  & 1.4904  & 1.8224  & 1.8230  & 1.8230  & $-0.5325$ \tabularnewline
4  & 2.8153  & 2.6450  & 2.0823  & 2.0823  & $-0.2687$ \tabularnewline
5  & 1.5101  & 2.1490  & 1.9215  & 1.9215  & 0.0070 \tabularnewline
6  & 3.2420  & 2.9600  & 2.2802  & 2.2802  & 0.2587 \tabularnewline
7  & 0.6163  & 0.1990  & 1.8797  & 1.8797  & 0.4434 \tabularnewline
8  &  &  & 2.4275  & 2.4275  & 0.5105 \tabularnewline
9  &  &  & 1.4964  & 0.7500  & 0.4026 \tabularnewline
10  &  &  &  &  & 0.0549 \tabularnewline
11  &  &  &  &  & $-1.6044$ \tabularnewline
\end{tabular}\end{ruledtabular} 
\end{table}

\end{document}